\documentclass[conference]{IEEEtran}
\IEEEoverridecommandlockouts
\usepackage{cite}
\usepackage{amsmath,amssymb,amsfonts}
\usepackage{algorithmic}
\usepackage{graphicx}
\usepackage{textcomp}
\usepackage{xcolor}
\usepackage{subcaption}
\usepackage{listings}
\def\BibTeX{{\rm B\kern-.05em{\sc i\kern-.025em b}\kern-.08em
    T\kern-.1667em\lower.7ex\hbox{E}\kern-.125emX}}
\usepackage{pifont}
\usepackage{booktabs}
\usepackage{hyperref}

\newcommand{\secref}[1]{Sec.~\ref{#1}}
\newcommand{\figref}[1]{Fig.~\ref{#1}}
\newcommand{\tabref}[1]{Table~\ref{#1}}

\newcommand{\cmark}{\ding{51}}
\newcommand{\xmark}{\ding{55}}

\lstdefinelanguage{json}{
    string=[s]{"}{"},
    stringstyle=\color{blue},
    comment=[l]{:},
    commentstyle=\color{black},
}

\begin{document}

\title{HyQDB: LLM-Assisted Debugging for Hybrid Quantum Workflows}

\author{
    \IEEEauthorblockN{Charlie Campbell}
    \IEEEauthorblockA{
        \textit{Department of Computing} \\
        \textit{Imperial College London}\\
        charlie.campbell22@ic.ac.uk
    }
    \and
    \IEEEauthorblockN{Hao Mark Chen}
    \IEEEauthorblockA{
        \textit{Department of Computing} \\
        \textit{Imperial College London}\\
        hao.chen20@imperial.ac.uk
    }
    \and
    \IEEEauthorblockN{Shuang Liang}
    \IEEEauthorblockA{
        \textit{Department of Computing} \\
        \textit{Imperial College London}\\
        shuang.liang@imperial.ac.uk
    }
    \and
    \IEEEauthorblockN{Hongxiang Fan}
    \IEEEauthorblockA{
        \textit{Department of Computing} \\
        \textit{Imperial College London}\\
        hongxiang.fan@imperial.ac.uk
    }
}

\maketitle

\begin{abstract}

% The majority of hybrid quantum program failures occur silently, with existing code generation and debugging tools having no mechanisms to catch this. 
Hybrid quantum program failures frequently occur silently, yet existing debugging tools provide limited support for detecting and repairing them. These faults dominate the failures reported by domain experts, yet existing tools evaluate on public data that under-represents this failure mode.
Our key insight is that faults divide into two classes that require different strategies: mechanical faults, which allow deterministic analysis, and conceptual faults, which require reconstructing the program's intent.  
To address this challenge,
we present HyQDB, a tiered agent that injects deterministic hardware, physics and optimization evidence into the LLM repair process.
When no evidence is detected, the agent treats this silence as a signal to escalate to a second intent-reconstruction tier, that infers the program's behaviour and reconciles it with the implementation. 
To evaluate HyQDB, we introduce QFaultBench, a benchmark built from an expert-derived fault taxonomy that represents the true failure modes of hybrid programs. On a held-out set of human-authored programs, HyQDB raises repair success over a standard LLM from 45\% to 75\%. We show that the escalation gate is crucial, giving a 20\% improvement in mechanical fault repair accuracy.
\end{abstract}

% \begin{IEEEkeywords}
% quantum design automation, hybrid quantum programs, AI design automation, quantum program debugging
% \end{IEEEkeywords}

\section{Introduction}

Hybrid quantum-classical programs interleave a quantum circuit with a classical optimization loop. This structure makes variational algorithms \cite{cerezo2021variational} tractable on noisy hardware \cite{preskill2018quantum}, but means a fault can occur at any of the boundaries without ever causing the program to crash. A circuit can be constructed correctly but parameterized incorrectly, causing the optimizer to unknowingly converge to the wrong value. Whereas a crash is self-announcing, a silent error risks corrupting the conclusions made from scientific experiments. This failure mode dominates expert-reported faults \cite{bensoussan2026taxonomy} and no current tool is able to detect or repair them.

A taxonomy of real hybrid faults found that over 40\% occur during parameterization and optimization, with over 50\% of faults concerning the interface between components~\cite{bensoussan2026taxonomy}. This study also found a split in fault symptoms. Only 35\% of GitHub faults produced a wrong output rather than a crash, whilst over 75\% of expert-reported faults followed this pattern.
This discrepancy is traced to why issues are reported. GitHub faults are filed when something visibly breaks, whereas experts reported not publishing many of the faults they encountered~\cite{bensoussan2026taxonomy}. This has an important consequence: any benchmark or tool built on public data is built from the reported failure mode, not the experienced one. This data bias means tools evaluated against public data are not evaluated against the failure modes that matter the most.

Furthermore, existing tooling for hybrid programs addresses only part of the fault space. Static analyzers, including LLM-based linters \cite{paltenghi2023lintq, cassieri2026beyond}, have no access to execution semantics and fail to resolve silent faults. LLM-based debuggers automate detection and repair, but rely on un-augmented judgment at each stage. QBugLM \cite{pham2026qbuglm} improves accuracy through iterative retry, with its taxonomy confined to syntactic and structural faults with circuits. MetaMorphQ \cite{nguyen2026metamorphq} derives physics-based invariants for variational circuits, but cannot reason about faults that manifest during training. Both of these solutions evaluate on canonical or synthetically generated circuits, not being tested on the real taxonomies shown in \cite{bensoussan2026taxonomy}.

\begin{figure}[t]
    \centering
    \vspace{-5mm}
    \includegraphics[width=\linewidth]{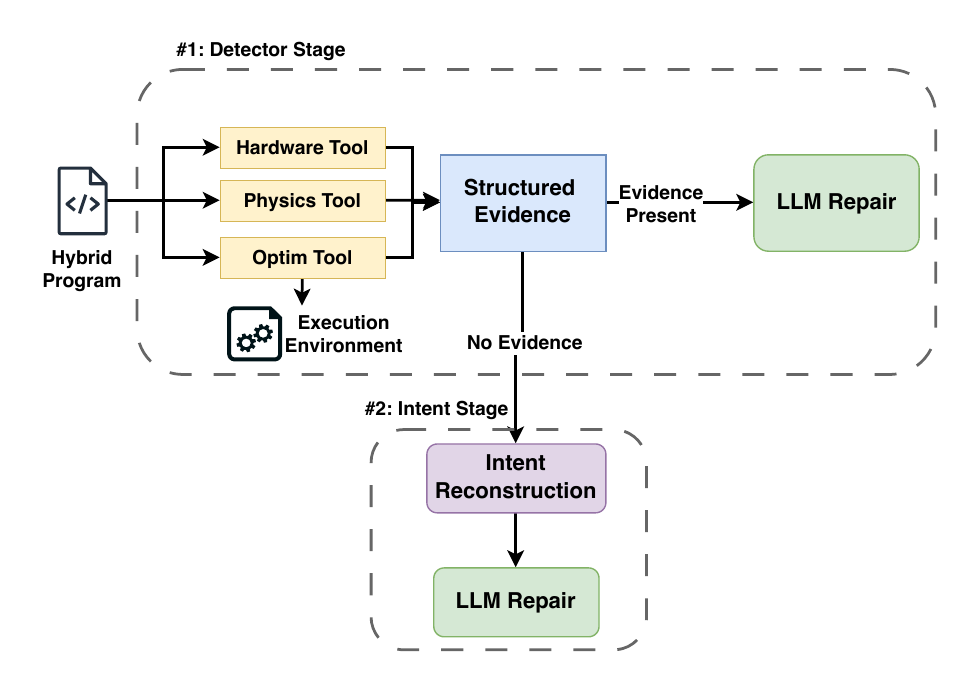}
    \caption{\textbf{HyQDB}: LLM-based hybrid program debugging. Static and runtime tools are used to identify deterministic faults. A gating mechanism is used to reconstruct intended semantic behavior to resolve faults that cannot be detected.}
    \label{fig:hyqdb}
\end{figure}

Existing methods apply a uniform strategy to every fault and do not diagnose fault type before attempting repair. We propose that hybrid faults separate into two key classes: mechanical faults resolved by static and runtime analysis, and conceptual faults resolved by reconstructing the program's intent. Based on this, we create HyQDB (\figref{fig:hyqdb}), a two-tier framework that uses deterministic analyzers to both classify fault types and inject evidence into an LLM to repair quantum programs. When no evidence is found, the framework routes to an intent-reconstruction tier designed to solve silent conceptual faults. We make the following key contributions:

\begin{enumerate}
    \item \textbf{HyQDB}, an agent with three deterministic analyzers (hardware, optimization and physics) that inject structured evidence into an LLM repair call.
    \item \textbf{QFaultBench}, a benchmark of 62 faults built from a known taxonomy \cite{bensoussan2026taxonomy}, verified with hidden execution-checked asserts. The benchmark is partitioned into a development set and a human-authored held-out set to demonstrate performance generalization.
    \item \textbf{A tiered escalation principle for hybrid fault repair}, in which the absence of a detector finding gates an intent-reconstruction step. Evaluated on QFaultBench, this signal correctly routes 76\% of faults to the appropriate repair strategy. 
\end{enumerate}

We make the code for both HyQDB and QFaultBench available on GitHub.\footnote{\url{https://github.com/CharlieC04/HyQDB}}

\section{Background and Motivation}

\begin{figure}[t]
    \centering
    \includegraphics[width=0.8\linewidth]{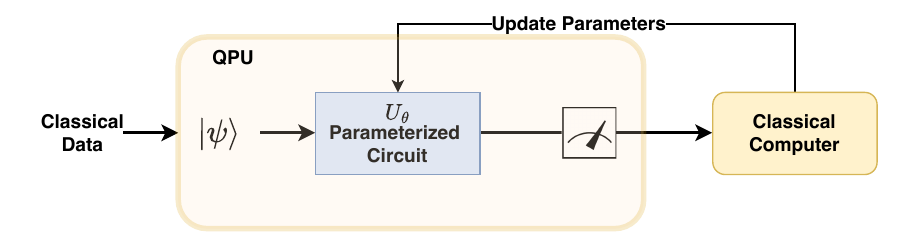}
    \caption{Outline of a hybrid quantum program.}
    \label{fig:hybrid_prog}
\end{figure}

Hybrid quantum programs are one of the leading near-term quantum applications, including estimating molecular ground states \cite{peruzzo2014variational}, solving optimization problems \cite{farhi2014quantum} and quantum machine learning \cite{havlivcek2019supervised}. A hybrid program (\figref{fig:hybrid_prog}) consists of two key phases. First a quantum circuit is parameterized and executed to produce a measurement, then a classical program consumes the output and updates the circuit's parameters. This loop repeats until the classical control script is satisfied the correct outcome has been reached.  Existing debugging tools focus entirely on the circuit, leaving the classical loop and its interface with the circuit - the region where most faults are concentrated - unsupported.

\subsection{Studies of Hybrid Quantum Software Faults}

The taxonomy in \cite{bensoussan2026taxonomy} is built from 133 Github faults with a further 52 faults from structured interviews with domain experts. Crucially, these two sources disagree on the primary symptom of faults. Only 35\% of Github faults produce a wrong output without a crash, whilst this number is 75\% in interviews. This finding means that any existing benchmark built upon publicly-mined data is not representative of the faults that real quantum developers face.

This is further supported by an independent survey of 26 developers \cite{zappin2025quantum}, which also found that an incorrect output without a crash is the most common type of bug. Importantly, this survey found that 69\% of developers use no dedicated debugging tool, suggesting current tools fail to match developer needs.

This data bias manifests in benchmarks such as Bugs4Q \cite{zhao2021bugs4q}, which collected bugs from public Github issues. Only 41\% of bugs identified followed the silent crash profile, with the rest related to issues with circuit design or development API. Further, a large portion of bugs found were not reproducible due to missing test cases or failure to run in the latest environment, demonstrating the low quality of publicly available data. 

Beyond taxonomy, the KCL study \cite{bensoussan2026taxonomy} found two current manual methods of debugging hybrid programs. The first was an optimization-first approach, in which gradient values are manually checked for non-convergence. The second is a problem-first approach, in which the Hamiltonian and qubit mapping are checked. Neither process is automated in current tooling, meaning both manual tracing and domain expertise are required to debug hybrid programs.

\begin{table}[h]
\centering
\begin{tabular}{@{}lccccc@{}}
\toprule
\textbf{Tool} & \textbf{Exec.} & \textbf{Determ.} & \textbf{Interface} & \textbf{Repair} & \textbf{Human} \\
& \textbf{evid.} & \textbf{/ Oracle-free} & \textbf{coverage} & & \textbf{eval.} \\
\midrule
LintQ~\cite{paltenghi2023lintq} & \xmark & \cmark & \xmark & \xmark & \cmark \\
LintQ-LLM~\cite{cassieri2026beyond} & \xmark & \cmark & \xmark & \xmark & \cmark \\
QBugLM~\cite{pham2026qbuglm} & \xmark & \xmark & \xmark & \cmark & \xmark \\
MetaMorphQ~\cite{nguyen2026metamorphq} & \xmark & \cmark & \xmark & \xmark & \xmark \\
\textbf{HyQDB (ours)} & \cmark & \cmark & \cmark & \cmark & \cmark \\
\bottomrule
\end{tabular}
\vspace{2mm}
\caption{Comparison of prior quantum-software debugging and verification tools against the capabilities required to address silent hybrid faults.}
\label{tab:related}
\end{table}

\subsection{Current Debugging Tools}

An overview of existing quantum debugging tools is shown in \tabref{tab:related}. Importantly, these tools are designed for debugging standard quantum circuits, not hybrid quantum programs.

Early quantum debugging tools used static analysis to detect faults with high precision \cite{paltenghi2023lintq}. However, due to the frequency at which quantum programming frameworks are updated, these tools become out of date very fast. Hence, they were extended to use LLMs to manage API updates whilst maintaining high precision. LintQ-LLM \cite{cassieri2026beyond} improves error detection precision from 62\% to 70\%, however performs poorly on complex data flow problems (30\%) and failed 10\% of tasks due to a limited context window. In addition, these tools are unable to resolve the errors that they detect.

Recent work expanded into tools that can automatically generate~\cite{campbell2025enhancing}, detect and repair bugs in quantum programs. QBugLM introduced a multi-agent framework to debug QASM code \cite{pham2026qbuglm}. The framework uses an LLM to create a structured bug report, which is then passed to another model to correct it. The key insight is that a single retry, using previous reports and attempted fixes, increases repair rate from 25\% to 80\%. However, this framework was evaluated on a limited number of LLM-augmented scripts that inject bugs based on a taxonomy collected from Github, which is known to not represent the true faults faced by quantum developers \cite{bensoussan2026taxonomy}.

\begin{figure*}[t]
    \centering

    \begin{subfigure}[c]{0.31\textwidth}
\begin{lstlisting}[language=Python]
theta = 0.1 * np.random.randn(*shape, requires_grad=True)
opt = qml.AdamOptimizer(0.1)

for step in range(150):
    opt.step(energy, theta) # BUG


\end{lstlisting}
        \caption{Buggy hybrid code}
        \label{subfig:code}
    \end{subfigure}
    \hfill
    \begin{subfigure}[c]{0.31\textwidth}
        \includegraphics[width=\linewidth]{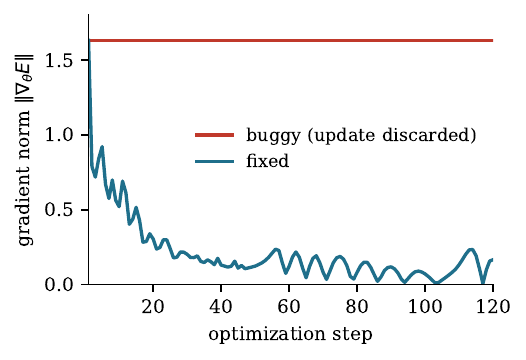}
        \caption{Gradient trajectory}
        \label{subfig:graph}
    \end{subfigure}
    \hfill 
    \begin{subfigure}[c]{0.31\textwidth}
\begin{lstlisting}[language=json]
{
    "qnode_executions": 60,
    "dynamic_status": "profiled",
    "output_trajectory_var": 0.0,
    "agent_insights": [
        "WARNING: The circuit output is frozen"
    ]
}
\end{lstlisting}
        \caption{Evidence JSON}
        \label{subfig:json}
    \end{subfigure}
    
    \caption{An example of a buggy hybrid program only detectable through runtime analysis. In (a), the parameters are not updated in the optimization loop. This causes the gradient in (b) to remain constant, which is detected in the tool output in (c).}
    \label{fig:optim_example}
\end{figure*}

Another approach to debugging is using physics-based invariants tested before optimization to validate whether variational quantum programs are correct. MetaMorphQ \cite{nguyen2026metamorphq} introduces a metamorphic testing framework for Variational Quantum Eigensolvers (VQE) \cite{peruzzo2014variational}. An example case involves negating a single Hamiltonian coefficient, which lets the optimizer reach a local minimum and results in convergence testing finding no error \cite{nguyen2026metamorphq}. The tool evaluates on a set of LLM-generated scripts, meaning that the performance of the framework on human-authored scripts remains unknown, as well as its generalizability to other circuits.

\subsection{Key Insights}

From analysis of existing methods, we learn three key insights.

\subsubsection{Insight \#1} Existing tools do not ground diagnosis in deterministic, execution-aware evidence. Linters \cite{paltenghi2023lintq, cassieri2026beyond} never execute the program, while QBugLM's \cite{pham2026qbuglm} decisions are made based on un-augmented LLM judgment. This motivates grounding decisions in data collected both before and during optimization, using deterministic information to make decisions.

\subsubsection{Insight \#2} Current methods cannot determine whether a fault is conceptual or mechanical. Circuits are checked before optimization begins, with no information about whether the program solves the desired problem. This motivates an intent-reconstruction step for faults that mechanical evidence cannot explain.

\subsubsection{Insight \#3} Every method discussed evaluates on data built from public Github data or LLM-generated circuits, which does not follow the fault taxonomy reported by experts \cite{bensoussan2026taxonomy}. This motivates the creation of a benchmark built from an expert-derived taxonomy that is validated against a set of held-out human-authored programs.

\section{HyQDB Design} \label{sec:design}

% This figure is not going to the correct page, should be top of page 3

\begin{figure*}[t]
    \centering
    \begin{subfigure}[c]{0.31\textwidth}
\begin{lstlisting}[language=Python]
@qml.qnode(dev)
def x_term(theta):
    var_ansatz(theta)
    return qml.expval(
            qml.PauliZ(0)) # BUG

\end{lstlisting}
        \caption{Buggy hybrid code}
        \label{subfig:code}
    \end{subfigure}
    \hfill
        \begin{subfigure}[c]{0.31\textwidth}
\begin{lstlisting}[language=json]
{
    "algorithm": "VQE energy estimation, H = X(0)",
    "expected_components": {
        "measurement": "x_term should measure PauliX(0)",
        "cost_function": "X expectation value"
    }
    "intent_violations": [
        "Function `x_term` measures Z basis"
    ]
}
\end{lstlisting}
        \caption{Intent evidence JSON}
        \label{subfig:json}
    \end{subfigure}
    \hfill 
\begin{subfigure}[c]{0.31\textwidth}
\begin{lstlisting}[language=Python]

@qml.qnode(dev)
def x_term(theta):
    var_ansatz(theta)
    return qml.expval(
            qml.PauliX(0))


\end{lstlisting}
        \caption{Fixed hybrid code}
        \label{subfig:code}
    \end{subfigure}
    
    \caption{An example of a buggy hybrid program that doesn't match detectable faults. The LLM detects the mismatch between basis measurement intent and resolves it.}
    \label{fig:intent_example}
\end{figure*}

HyQDB uses pre-computed evidence injection to deterministically provide an LLM with evidence about faults in a program. Earlier iterations used a ReAct \cite{yao2022react} agentic loop, giving the model tool calling access to each analyzer. However, the model often called the wrong tool or hallucinated results, motivating the switch to evidence injection. This is supported by QBugLM's finding that single-shot prompting outperformed ReAct, contrary to the expectation that explicit reasoning scaffolds would improve performance \cite{pham2026qbuglm}.

For each script, all three analyzers collect any potential evidence as a structured JSON. If any potential faults are found, this is injected into a call to an LLM that attempts to correct the code. In some cases, the tools can point to a specific location in the source code that causes a fault, whilst in others the model must infer the location from the provided evidence.

\subsection{Deterministic Analyzers}

The three analyzers automate the two manual debugging procedures reported in \cite{bensoussan2026taxonomy}. Each analyzer collects structured evidence that is passed to the agent at repair time:

\subsubsection{\textbf{Hardware Detector}} This analyzer extracts hardware properties such as the target device, gate set and entangling-gate usage by scanning the source code text. It can detect two concrete faults. First, a native-gate mismatch between an implied target hardware and present gates (e.g. a trapped ion device using \texttt{CNOT} gates). Second, a mixed gate-set when both \texttt{CNOT} and Ising-family gates are used in the same circuit.

\subsubsection{\textbf{Optimization Detector}} Optimization faults are the largest category in \cite{bensoussan2026taxonomy} and are often invisible to static analysis. The manual tracing procedures reported by practitioners exist because such faults only manifest themselves at runtime. This analyzer combines a static abstract syntax tree (AST) scan with dynamic profiling. The script is executed in a PennyLane \cite{bergholm2022pennylaneautomaticdifferentiationhybrid} environment, with every \texttt{QNode.\_\_call\_\_} intercepted to allow the tool to view trajectory during training.

The static scan performs three checks. First, trainable parameters initialized to zero are flagged since they place the optimizer at a symmetric point with identical gradients. Second, any defined learning rate is checked against the bounds $\textit{lr}\in [10^{-2}, 2.0 ]$, which were calibrated on the development set of QFaultBench. Third, ansatz depth is checked against a threshold associated with poor convergence, similarly calibrated using QFaultBench.

Following this, the script is executed and the variance of the output trajectory is examined. Minimal variance indicates frozen optimization, which can be caused by issues with the classical library interface or poor initial parameter values. In contrast, large variance indicates divergence and could be caused by poor optimization parameter values.

\figref{fig:optim_example} shows an example of a fault that is only detectable via runtime analysis: the program does not correctly update the value of the parameter \texttt{theta} using the output of the \texttt{opt.step} function, which causes the circuit to be evaluated at the same point in every iteration. Whilst a static check would be unable to consistently check this behavior, the zero variance found during the runtime analysis can be used to provide evidence of a parameter update fault.

\subsubsection{\textbf{Physics Detector}} This analyzer infers the problem domain from imports, state-preparation operators and ansatz templates. It then applies domain specific checks to the problem type to ensure it matches known invariants. The current focus of the tool is quantum chemistry, with the tool checking for the presence of operators known to break quantities such as particle number or spin symmetry, for example no \texttt{BasisState} preparation or the presence of \texttt{StronglyEntanglingLayers}.  If no faults are found, the predicted domain can be passed to the intent-reconstruction tier to aid inference.

The tool additionally performs general checks, such as the absence of a measurement and low shot count. These checks verify that the script does not violate properties of reliable quantum computation, rather than checking whether the script solves the intended problem.

\subsection{Intent Reconstruction}

The three analyzers check a fixed set of patterns, designed to match the most prevalent faults in current hybrid programs. However, a lack of evidence does not necessarily mean that the program is correct, but that the potential fault may be outside of the detectable set. In QFaultBench, such faults map to the \textit{conceptual} category in \cite{bensoussan2026taxonomy}, in which there is a mismatch between the defined program and the intended behavior. Importantly, this fault has zero instances in public GitHub data but is the largest category raised in expert interviews, making it an important, and as yet unresolved, problem.

When the analyzers find no faults, HyQDB escalates to a bounded intent-reconstruction tier, rather than attempting an evidence-free repair. The first call attempts to reconstruct the program's intended behavior, inferring intent from function and variable names, comments and problem setup. It returns a predicted algorithm class, the expected outcome of a correct implementation and points to where the implementation differs from the intent. The second call is given this specification and asked to reconcile the implementation against it. Importantly, if the first call fails to infer an intent, the tier reverts to a standard repair attempt, ensuring performance does not degrade against a standard LLM.

\figref{fig:intent_example} shows an example of how the intent reconstruction tier works. The function defined measures in the Z-basis, however the LLM is able to infer from the function name that the intent is to carry out a measurement in the X-basis. Using this information, it is then able to update the function to carry out the correct behavior. This fault is not possible to resolve via static detectors as it requires natural language understanding of what the author was trying to achieve based on the defined program.

\section{QFaultBench}

To evaluate HyQDB, we created QFaultBench, a benchmark that reflects the dominant failure modes that real hybrid developers face \cite{bensoussan2026taxonomy}. The benchmark consists of 62 repair tasks spanning ten real PennyLane workloads, in which each task executes without raising an exception but contains a single injected fault that silently causes it to break.

\subsection{Construction}

Each item is produced using a four stage pipeline. Base programs are drawn from working PennyLane \cite{bergholm2022pennylaneautomaticdifferentiationhybrid} demos, which are verified to converge correctly in the code environment. Then, a modified script is created for each fault type that injects a localized edit into the script that silently breaks it. For the development set, these alterations were made using an LLM (gemini-flash-3.6). For the held-out set, these alterations were made by a human expert. Each variant is then ran and validated, only being admitted to the benchmark if it fails the test's hidden assert. Finally, the script is cleaned using \texttt{ast.unparse} to remove all comments and help reduce the chance of the model guessing the fault from hints in the code.

\subsection{Fault Taxonomy}

\begin{table*}[t]
\centering
\caption{Fault taxonomy. Dev/held-out columns give the number of benchmark items per family in each split.}
\label{tab:taxonomy}
\begin{tabular}{@{}lccp{7cm}@{}}
\toprule
\textbf{Family} & \textbf{Dev} & \textbf{Held-out} & \textbf{Focus} \\
\midrule
Conceptual & 10 & 8 & Runs but implements the wrong algorithm (cost sign, missing mixer/entangler, wrong basis, missing re-upload \cite{perez2020data}) \\
\addlinespace
Initialization / optimization & 11 & 8 & Trainability failures: symmetric saddle init, barren plateaus \cite{mcclean2018barren}, divergent / vanishing step size \\
\addlinespace
Interface / differentiability & 8 & 6 & The classical-quantum interface: severed autograd chains, tensor-shape mismatches \\
\addlinespace
Measurement / statistics & 3 & 3 & Estimator quality: too few measurement shots (shot noise) \\
\addlinespace
Hardware / compilation & 2 & 3 & Device realizability: native-gate / topology violations, excessive two-qubit depth \\
\bottomrule
\end{tabular}
\end{table*}

The fault taxonomy groups 13 fault types defined in \cite{bensoussan2026taxonomy} into five families. This taxonomy was fixed before HyQDB was designed, ensuring independence from the framework. The chosen faults are unique to hybrid quantum programs, focusing on silent semantic faults and the interaction between the quantum and classical parts of the program.

Of the 62 tasks, 34 form the development set and 28 form the held-out set. The base scripts used for each set were kept independent to ensure no leakage between framework development and evaluation. Programs are kept short (less than 100 lines) and use either \texttt{PennyLane-autograd} \cite{bergholm2022pennylaneautomaticdifferentiationhybrid} or \texttt{JAX} \cite{jax2018github} as the classical optimization framework.

\tabref{tab:taxonomy} shows the distribution of the fault types across the two benchmark sets. The key focus of the benchmark is on conceptual and interface faults. A limited number of examples from other areas are included in the benchmark to ensure the main fault types found are fully covered.

\subsection{Benchmark Fairness}

Three safeguards were implemented to prevent a model from passing tests due to recall rather than reasoning. First, comment stripping removes every injection marker and natural language hint, meaning the model only sees the runnable buggy program. Second, exact-value oracles were excluded from the benchmark to ensure the LLM must solve the problem instead of deriving the solution. Finally, the held-out split is entirely human authored, meaning the LLM has no memorization of the base scripts used. 

\section{Evaluation} \label{sec:eval}

We evaluate HyQDB across three configurations. The first, a baseline un-augmented model (\texttt{deepseek-v4-flash} \cite{deepseekai2026deepseekv4}), which we also use for all LLM calls within the framework; the second includes the detector analysis stage; the third, the full two-tier HyQDB framework. All results are reported as mean and standard deviation over three runs, with variance occurring due to different LLM reasoning. 

\subsection{Key Results}

\begin{figure}[h]
    \centering
    \includegraphics[width=0.9\linewidth]{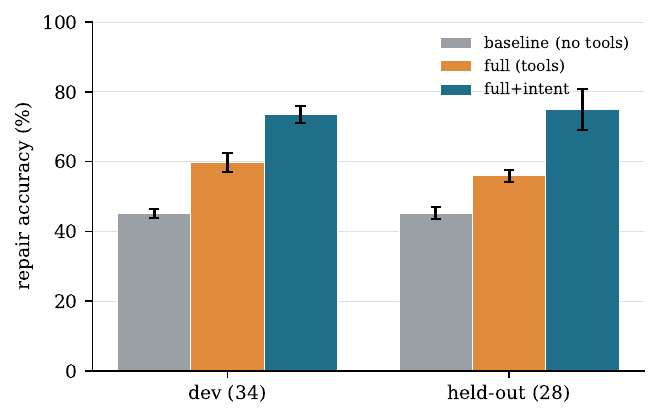}
    \caption{Accuracy of HyQDB on development and held-out sets of QFaultBench.}
    \label{fig:accuracy}
\end{figure}

We first demonstrate that tool use improves repair accuracy on both the development and held-out set. \figref{fig:accuracy} shows that repair accuracy improves from 45±1.4\% to 60±2.8\% when deterministic analyzers are used, rising to 74±2.4\% when the intent-reconstruction tier is added. Importantly, this behavior is mirrored on the held-out set, with the base model achieving 45±1.7\% repair accuracy, rising to 56±1.7\% and 75±5.8\% respectively. The monotonic improvement on the held-out set demonstrates that the framework is able to generalize to unseen problems and is able to resolve silent errors in real human-authored quantum programs. The baseline model performs identically on both sets, implying that they are both of similar difficulty.

\begin{figure}[h]
    \centering
    \includegraphics[width=0.9\linewidth]{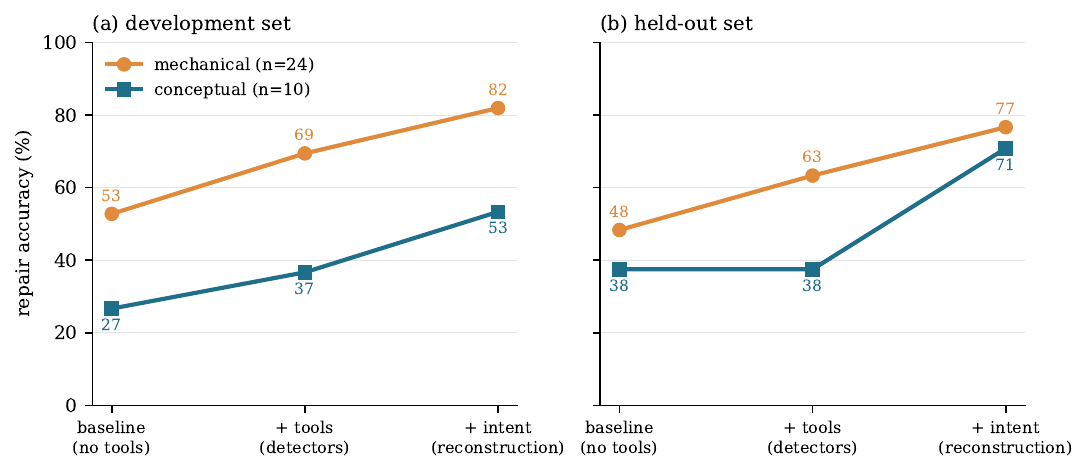}
    \caption{Repair accuracy per fault type on each QFaultBench set.}
    \label{fig:fault_type_acc}
\end{figure}

\figref{fig:fault_type_acc} isolates the two key fault classes and provides the evidence for the two-tier design. On both splits, mechanical faults improve when detector evidence is added (53\% to 69\% and 48\% to 63\%). However, conceptual faults remain at similar levels when this is added, only increasing once the intent-reconstruction tier is added. This result justifies the escalation gate, rather than always running both tiers. The improvement of mechanical faults on the intent-reconstruction tier is an artifact of the detectors failing to find any evidence, discussed further in \secref{subsec:routing}.

\begin{figure}[h]
    \centering
    \includegraphics[width=0.7\linewidth]{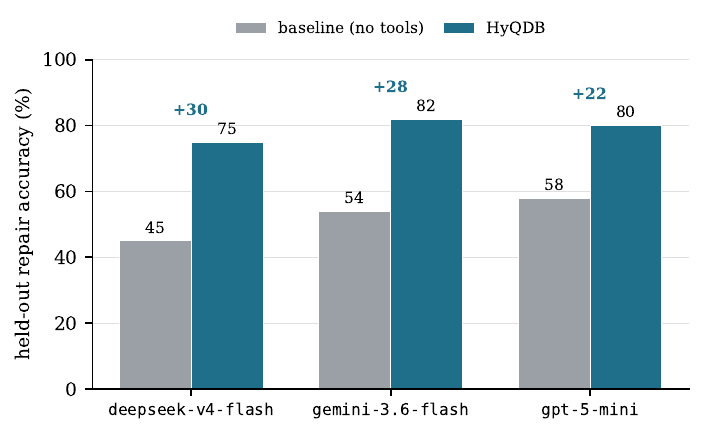}
    \caption{Performance of HyQDB on various models.}
    \label{fig:models}
\end{figure}

To demonstrate that the performance gains of HyQDB arise from the framework design, rather than being an artifact of the model, we test across three base LLMs. \figref{fig:models} demonstrates that our framework improves performance across all models, reaching a similar final repair accuracy in all cases. This suggests that the current accuracy is a limitation of the framework, rather than the model itself, motivating the development of additional detectors and intent-reconstruction tools.

\subsection{Detailed Fault Analysis}

\begin{figure}[h]
    \centering
    \includegraphics[width=0.9\linewidth]{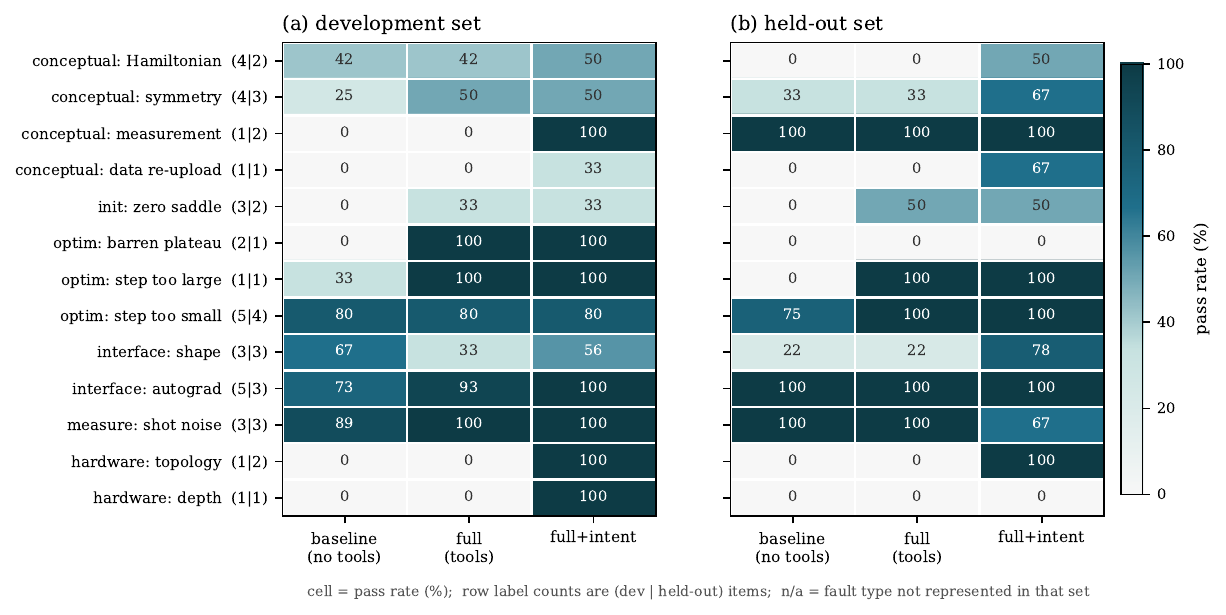}
    \caption{A per-category breakdown of HyQDB's ability to correct silent faults in hybrid programs.}
    \label{fig:category_accuracy}
\end{figure}

\begin{figure}[h]
    \centering
    \includegraphics[width=0.9\linewidth]{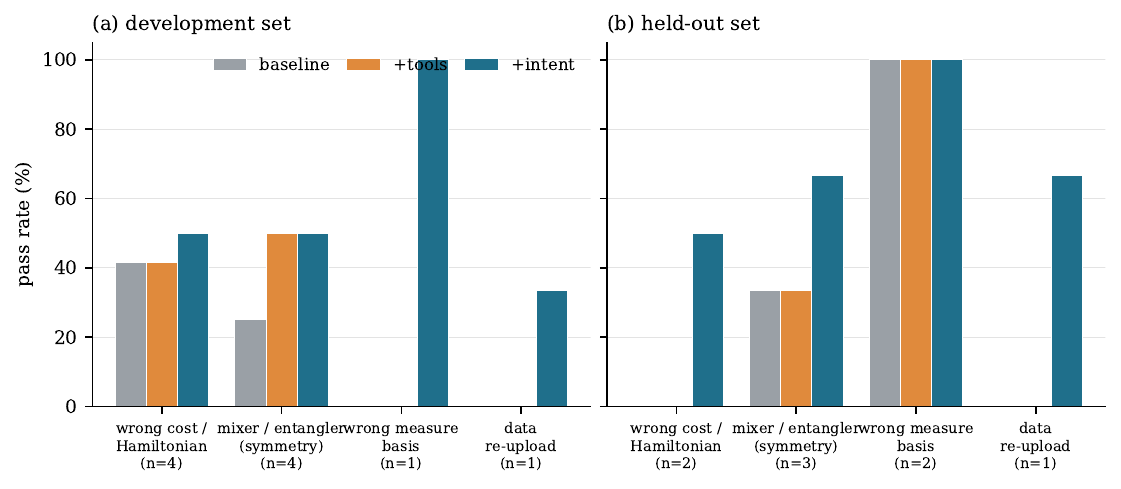}
    \caption{A per-fault breakdown of \texttt{conceptual} category bugs.}
    \label{fig:conceptual_accuracy}
\end{figure}

The architecture in \secref{sec:design} is designed such that faults with a fixed, checkable signature are resolved via evidence injection, whereas faults with no signature are resolved via intent-reconstruction. We demonstrate this principle by testing the repair rate of each bug category in QFaultBench, shown in \figref{fig:category_accuracy} and \figref{fig:conceptual_accuracy}.

\figref{fig:category_accuracy} demonstrates that detectable categories such as \texttt{optimization} and \texttt{interface} are improved directly by injecting evidence into the model. The majority of these faults improve between the baseline and evidence injection, with no further improvement once intent-reconstruction is applied. This demonstrates that evidence is found for the majority of these faults, even if the tier cannot resolve the problem.

The reverse pattern is present in the \texttt{conceptual} category, with improvements usually only arising once intent-reconstruction is applied. This is more pronounced on the held-out set, with the LLM unable to solve the conceptual problems via recall without augmentation due to the more complex nature of the problems. There are some problems that the model is unable to solve without augmentation, which is evidence of the model's base reasoning ability. However, providing detector evidence does not improve accuracy on these tasks, demonstrating that semantic understanding is needed to resolve them, rather than static or runtime analysis. This is further demonstrated in \figref{fig:conceptual_accuracy}, which demonstrates that intent-reconstruction achieves higher repair accuracy on the majority of task categories.

There are two results that do not follow the general pattern. The first is \textit{barren\_plateau}, which is fully resolved on the development set, but not on the held-out set. This is likely due to the increased difficulty of the held-out problem, in which the barren plateau is successfully detected via low variance, but the model unable to resolve it. The second is the \textit{depth} bug, which only resolves under intent-reconstruction despite not being a conceptual fault. This is likely due to detector failure, with the tool not fully designed to solve problems in the \texttt{hardware} category due to their low occurrence. This is an open problem that needs to be resolved in future iterations of the framework. 

Importantly, intent-reconstruction is not uniformly effective across conceptual faults. The method can reliably repair faults where the component is \textit{present but wrong}, such as a measurement in an incorrect basis. This is because the intended behavior is already there, and a localized edit can be used to resolve it. The method is weak when a required component is missing, such as in the data re-uploading case. This splits conceptual errors into two key groups and identifies the most difficult type of repair task: reconstructing missing structure, rather than resolving incorrect structure.

\subsection{Routing Mechanism} \label{subsec:routing}

\begin{figure}
    \centering
    \includegraphics[width=0.9\linewidth]{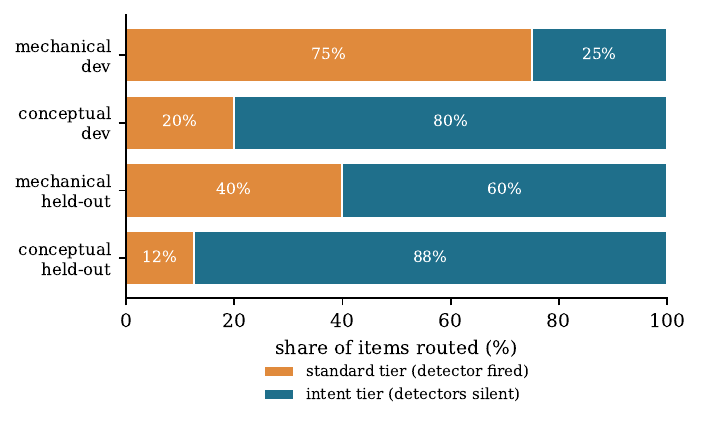}
    \caption{Breakdown of which faults get directed to each tier.}
    \label{fig:route_share}
\end{figure}

\begin{figure}[h]
    \centering
    \includegraphics[width=0.9\linewidth]{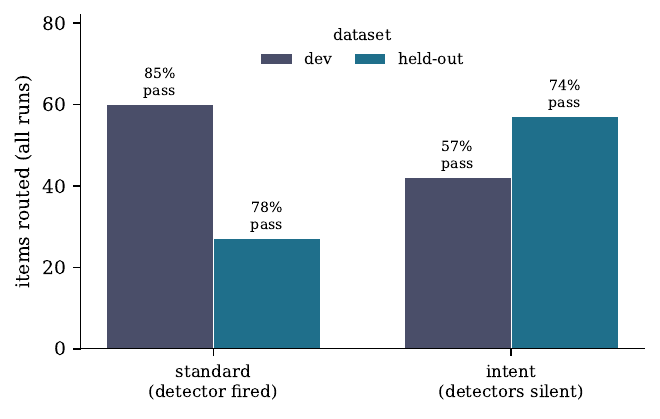}
    \caption{Accuracy of routing mechanism against the fault type.}
    \label{fig:route_accuracy}
\end{figure}

The primary principle of the framework is that detector silence indicates a conceptual fault. \figref{fig:route_accuracy} reports pass rate by tier and \figref{fig:route_share} reports the composition of each tier, demonstrating the accuracy of the escalation mechanism.

\figref{fig:route_share} shows that detector silence is an appropriate escalation mechanism for conceptual faults. On the development set, the framework finds evidence for 75\% of mechanical faults, whereas 80\% of conceptual faults are correctly routed to the intent tier. On the held-out set, 88\% of conceptual faults are correctly routed, demonstrating that the detectors are correctly finding the information they are built to, with out-of-class faults correctly routing to the intent tier. 

However, on the held-out set, detectors are silent for 60\% of mechanical faults. There are two key contributors to this. The first is that held-out tasks are inherently more difficult, as they are human-authored rather than being constructed from PennyLane examples \cite{bergholm2022pennylaneautomaticdifferentiationhybrid}. This means they are less likely to contain the standard faults that the detectors are designed to find. The second is that the detectors were tuned on the characteristics found in the development set. For example, the learning rate is gated on values $10^{-2}$ and $2.0$ based on observed values. However, a rate of $1.5$ may lead to poor convergence on problems in the held-out set, but would be classified by the detector as correct. This means that some problems are incorrectly routed to the intent tier, causing the framework to do more work than needed whilst still repairing the faults.

\figref{fig:route_accuracy} demonstrates that the accuracy of the detector tier is consistent across both benchmark sets (85\% vs 78\%), although the held-out set contains less faults that get routed to this category. The accuracy of the intent tier is much lower on the development set (57\% vs 74\%), which is likely another artifact of the erroneous detector silence on the held-out set. Since more mechanical faults are routed to this tier on the held-out set, the repair rate is higher as conceptual faults are typically harder to resolve. However, this demonstrates that the framework is still correctly resolving the mechanical faults even with the lack of evidence by using the intent reconstruction mechanism.

\begin{table}[h]
\caption{Effect of the escalation gate. \emph{gated} runs intent
reconstruction only when detectors are silent; \emph{no gate} runs it on
every task.}
\label{tab:gate}
\centering
\footnotesize
\begin{tabular}{lccc@{\hskip 10pt}ccc}
\toprule
& \multicolumn{3}{c}{Development} & \multicolumn{3}{c}{Held-out}\\
\cmidrule(lr){2-4}\cmidrule(lr){5-7}
Configuration & Mech & Conc & All & Mech & Conc & All\\
\midrule
full (detectors)         & 69 & 37 & 60 & 63 & 38 & 56\\
full+intent (no gate)    & 62 & 60 & 62 & 58 & 67 & 61\\
full+intent (gated)      & \textbf{82} & 53 & \textbf{74} & \textbf{77} & 71 & \textbf{75}\\
\bottomrule
\end{tabular}
\end{table}

A key design choice of the framework is that intent-reconstruction is only invoked when the detectors are silent. We test this mechanism by removing the gate and escalating all programs to the intent tier.

\tabref{tab:gate} demonstrates that removing the gate reduces the repair accuracy for mechanical faults by ~19\% on both sets. Running intent reconstruction on these faults discards the deterministic fault localization carried out by the detectors, making the repair job more complex. The gating mechanism does not change the performance of conceptual faults, which do not receive detector evidence and route to the intent tier under either policy.

\subsection{Token Usage}

\begin{figure}[t]
    \centering
    \includegraphics[width=0.8\linewidth]{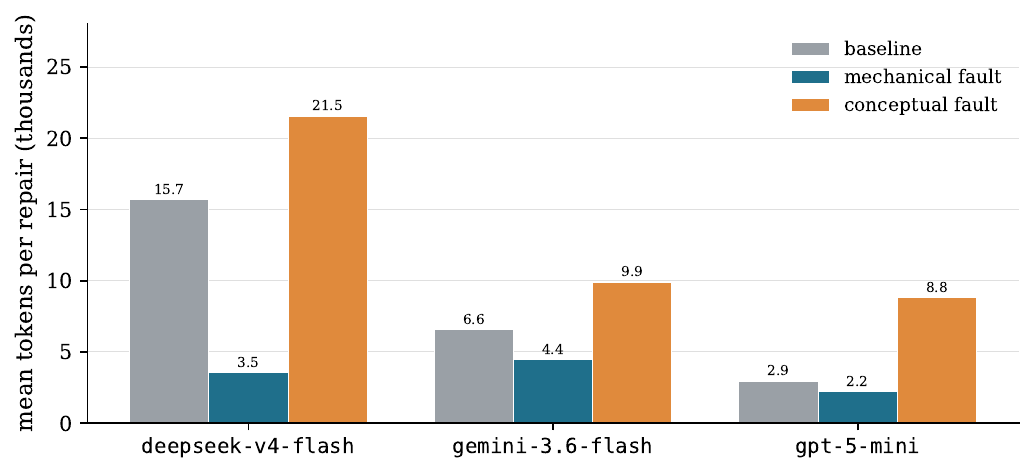}
    \caption{Token cost per fault class for each model tested.}
    \label{fig:tokens}
\end{figure}

Since each tier uses different model calls, we measure token cost separately for each fault class, shown in \figref{fig:tokens}. We see that mechanical faults use less tokens than an unaugmented baseline repair. This is because the detector gives a localized diagnostic, meaning the model is able to provide a direct solution. In the baseline, the model often fails to produce a valid code output on the first try, requiring a retry. A conceptual fault costs the most, since it requires an extra call to reconstruct the program's intent before the repair call. This causes the token count to increase by 2-4× relative to a detector fault.

The difference in token counts between models is a reflection of their design. For example \texttt{deepseek-v4-flash} often emits long reasoning chains and exceeds generation limits, requiring retries.

\section{Limitations \& Discussion}

Our framework has three key limitations. First, QFaultBench contains only 62 tasks, with some fault families represented by only a few examples. This means the per-category results are only indicative of expected performance. Second, the detector thresholds are calibrated on the development set and do not perfectly transfer to the held-out programs, demonstrated by the erroneous detector silence. Finally, the detectors present cover only a bounded set of patterns, with other mechanical faults incorrectly escalating to the intent-reconstruction tier. The multi model results, which all converge to a single final repair rate, suggests that the performance ceiling is an artifact of detector coverage rather than the reasoning ability of the model.

\section{Future Work}

The analysis in \secref{sec:eval} reveals three key directions for future work. First, intent-reconstruction corrects components that are present but wrong more reliably than it restores missing components. Extending this tier to propose and verify absent structures is the key next step. Second, the generalization gap for mechanical fault detection stems from fixed thresholds on detectors tuned on the development set of QFaultBench. Replacing these with algorithm-aware calibration would reduce the erroneous error detection seen on the held-out set. Third, the \texttt{hardware} fault category is currently underserved by existing tools. Creating a compilation-aware analyzer that reasons about device topology and SWAP-count~\cite{fan2022optimizing} would prevent these faults from being raised to the intent-reconstruction tier, which is not designed to solve these kinds of problems.

Additional work on QFaultBench should explore expanding the benchmark to more types of algorithms and fault-tolerant quantum workflows~\cite{liang2026coset, ziad2026greenpeas}, whilst also increasing the number of occurrences of each fault type in the benchmark.

\section{Conclusion}

This paper addresses the silent faults that dominate real hybrid fault taxonomies \cite{bensoussan2026taxonomy} yet remain undetected and unresolved by existing tooling. We introduce HyQDB, a two-tier agent that resolves each fault by class, applying deterministic evidence injection to mechanical faults and intent-reconstruction to conceptual faults. We also introduce QFaultBench, a benchmark built from an expert-derived taxonomy and validated with hidden asserts. We include a human-authored held out set to ensure tool generalizability and test performance on real-world problems. Our central results demonstrate that detector evidence lifts mechanical faults, whilst semantic faults require intent-reconstruction. On the held out set of human-authored programs, HyQDB improves repair accuracy over a base LLM from 45\% to 75\%, demonstrating that the approach generalizes to unseen problems. By grounding repair in execution evidence and semantic context, HyQDB detects and resolves the silent faults that threaten the correctness of hybrid quantum results, removing the need for complex manual debugging.

\bibliographystyle{IEEEtranS}
\bibliography{ref}

\end{document}